\documentclass[conference]{IEEEtran}
\IEEEoverridecommandlockouts
\usepackage{amsmath,amssymb,amsfonts}
\usepackage{graphicx}
\usepackage{textcomp}
\usepackage{xcolor}
\usepackage[numbers]{natbib}
\usepackage{stackengine}
\usepackage{multirow}
\usepackage{academicons}
\usepackage{microtype}
\usepackage{natbib}
\usepackage[T1]{fontenc}
\usepackage{graphicx}
\usepackage{tabularx}
\usepackage{hyphenat} 
\usepackage{subfig}
\usepackage{float}
\usepackage[linesnumbered,ruled,lined]{algorithm2e} 
\usepackage{algpseudocode}
\definecolor{orcidlogocol}{HTML}{A6CE39}

\def\BibTeX{{\rm B\kern-.05em{\sc i\kern-.025em b}\kern-.08em
    T\kern-.1667em\lower.7ex\hbox{E}\kern-.125emX}}

\begin{document}

\title{A Framework for Spatio-Temporal Graph Analytics In Field Sports}

\author{
\IEEEauthorblockN{Valerio Antonini\IEEEauthorrefmark{1}, Michael Scriney\IEEEauthorrefmark{2}, Alessandra Mileo\IEEEauthorrefmark{2}, and Mark Roantree \IEEEauthorrefmark{2} }

\IEEEauthorblockA{\IEEEauthorrefmark{1} School of Computing, Dublin City University, Dublin, Ireland\\ Email: valerio.antonini3@mail.dcu.ie}

\IEEEauthorblockA{\IEEEauthorrefmark{2} Insight Centre for Data Analytics, Dublin City University, Dublin, Ireland\\ Email: \{michael.scriney, alessandra.mileo, mark.roantree\}@dcu.ie}
}

\maketitle

\begin{abstract}
The global sports analytics industry has a market value of USD 3.78 billion in 2023. The increase of wearables such as GPS sensors has provided analysts with large fine-grained datasets detailing player performance. Traditional analysis of this data focuses on individual athletes with measures of internal and external loading such as distance covered in speed zones or rate of perceived exertion. However these metrics do not provide enough information to understand team dynamics within field sports. The spatio-temporal nature of match play necessitates an investment in date-engineering to adequately transform the data into a suitable format to extract features such as areas of activity. In this paper we present an approach to construct Time-Window Spatial Activity Graphs (TWGs) for field sports. Using GPS data obtained from Gaelic Football matches we demonstrate how our approach can be utilised to extract spatio-temporal features from GPS sensor data. 
\end{abstract}


\section{Introduction}

Spatio-temporal data refers to data that has both spatial (geographical) and temporal (time-related) components, exhibiting variations in both space and time. This type of data is commonly encountered in various fields such as environmental science \cite{wu2021}, agriculture \cite{MMOR2017}, transportation \cite{zhao2019}, \cite{zheng2020}, \cite{CNCR2024}, epidemiology \cite{jiang2020}, and meteorology \cite{ni2022}, among others. Over the last decade, spatio-temporal data have been hugely involved in sport thanks to the introduction of the use of sensors and in particular Global Positioning System (GPS). GPS sensors record a fixed number of observations (10 or 100 depending on the device) for each second of the variables latitude, longitude, speed (meters/seconds), and acceleration (meters$^{2}$/seconds). Sport organizations are investing in tracking systems to gain competitive advantages through harnessing recorded data, with the aim of making more data-driven decisions \cite{torres2022}. GPS data have emerged as a valuable resource for researchers and sports professionals to identify speed zones \cite{coutts2010} \cite{CMR2008}, define athletes' external load during competition \cite{malone2016}, identify sequential and recurrent players movements \cite{white2022}, among other applications. Recently, machine learning approaches, such as ensemble learning \cite{sheridan2024}, graph clustering \cite{brady2021}, and neural-network based \cite{kim2022}, have been applied to extract new insights. Thus far, GPS analytics focused on the analysis of individual players, to improve physical performances or prevent injuries, but have neglected the examination of team dynamics. More complex analyses such as understanding the relationship between players during game time, in terms of direction and intensity of movement is not currently addressed. The players’ profiles have been investigated without considering the interactions between teammates. To address the need for enriched individual and collective physical metrics, it is required the analysis of a network comprising players, locations and movements, and to understand how that network evolves over time. Spatio-temporal data exhibit interconnected dependencies within both temporal and spatial dimensions, deviating from the assumption of independence or identical distribution for individual instances~\cite{ferreira2020}. Due to the spatial and temporal components, GPS data are suitable for spatio-temporal graph representation. A spatio-temporal graph is a type of graph where usually nodes represent locations and edges represent events that start in a location and end in a different one. 

\textbf{Contribution.}
In this work, we introduce a graph-based framework to represent the journey of Gaelic Football (GF) players of a team during games and extract patterns and insights about players' movements. We are interested in analyzing the areas in the pitch with higher levels of action, making comparisons across games and players. Utilizing spatio-temporal graph representations can help sports scientists and coaches in making data-driven decisions based on objective insights derived from real-time player tracking data. This approach enhances the accuracy and effectiveness of performance analysis, player evaluation, and strategic planning, ultimately contributing to better team performance outcomes.

\textbf{Paper Structure.}
The remainder of this paper is structured as follows:
Section \ref{sec:rel_res} introduces past research in spatio-temporal graphs; Section \ref{sec:method} describes the required steps to realize the spatio-temporal framework; Section \ref{sec:results} presents the results obtained by the framework on a case study; Conclusion discussion and future work are presented in Section \ref{sec:conc}.

\section{Related Research}
\label{sec:rel_res}
Spatio-temporal graphs offer a powerful framework for analyzing GPS data, enabling the representation of both spatial locations and temporal dynamics over time. By incorporating both spatial and temporal dimensions, these graphs facilitate the exploration of movement patterns, trajectory analysis, and the understanding of dynamic interactions in various applications.

Authors in~\cite{vega-oliveros2019} present two different graph models: a static and a spatio-temporal graph, to analyze the fire activity in the Amazon basin (years 2003-2018). The static graph represent the fire activity for the entire period, while the spatio-temporal graph is based on the same building-rules but represent one week of activity. The area of interest is divided into a grid. Each cell grid represent a node of the network. Two nodes are connected when two successive events in chronological order happen in the grid cells corresponding to the nodes. This spatio-temporal graph representation permits to analyze the areas most affected by fires (centrality score) and the areas of similar activity (community-detection) over the different windows of time. In a successive work, same authors~\cite{ferreira2020} propose Chronnet, a chronological network-based model for analyzing spatio-temporal data. In Chronnet, the nodes are represented by areas of same size obtained by dividing a geometric space into grid cells. The areas are connected when consecutive chronological events happen in the two distinct cells. The graph model has been tested on three artificial datasets and a real dataset describing the global fire activity across the years 2003-2018. The authors demonstrated the capability of the suggested graph-based model to extract data characteristics that extend beyond basic statistics, such as frequent patterns, spatial variations, and spatio-temporal clusters. A frequency-based spatio temporal graph network is proposed in~\cite{jia2023} to analyze and compare bike-sharing mobility patterns in different years, before (2019), during (2020), and after (2021 and 2022) Covid-19 outbreak. In this study, the nodes are the location in the area of interest and weighted edges represent the frequency of trips starting in a location and ending in a different one. The metrics involved in analyzing the trend of mobility patterns are nodes' connectivity, clustering coefficient, and modularity-based community detection. The application of graphs in sports analytics provides a comprehensive framework for analyzing complex interactions and patterns among players, teams, and game elements. By representing players or areas of the pitch as nodes and actions as edges, graphs allow for insightful analyses of team strategies, player dynamics, and tactical decision-making, offering valuable insights for improving performances and enhancing data-driven coaching strategies.
In their study, \cite{gama2014} implemented betweenness centrality scores to identify the principal players during attacking actions in football. They represent the players as nodes and compute the number of interactions (performed or received passes) to define weighted-directed edges among them. In our study, we applied similar metrics to study the importance of nodes in the graph. Our research diverges in terms of graph topology, where we represent areas of the pitch as nodes rather than players, and movements at different speed zones as edges, contrasting with passing between teammates. Furthermore, our research objectives vary as we focus on assessing the significance of specific areas on the pitch rather than the importance of players.
Researchers in \cite{silva2017}, proposed the uPATO software, which offers the capability to generate adjacency matrices that illustrate the passing interactions among players within a team throughout a match or across different phases of a match. The main functionality provided by uPATO is the possibility to asses team performance over distinct time intervals within a match. Even in this case, our research diverges in both purpose and graph topology, although common aspects such as the use of centrality scores.
Passing networks are analysed also in \cite{goncalves2017}. The authors explored how passing networks and positioning variables (involved as attributes of the players-nodes) are linked to the outcome of elite young football matches. In particular, they found that lower betweenness centrality scores (describing passing dependency to given players) and high closeness centrality scores (describing intra-team well-connected passing relations) are related to better outcomes. In this study, although the authors used a two-step cluster analysis to classify teammates' positioning, they didn't delve deeply into the spatial aspect. They mainly focused on static positions on the field without looking into the dynamic movements of players, which could provide interesting insights into how players interact.
In a recent football study \cite{raabe2023}, a spatio-temporal graph was constructed with players as nodes and weighted edges representing distances between them on the pitch, including temporal connections between the same player across consecutive seconds. Various player attributes were incorporated, and a 15-second segment was analyzed for binary classification of possession change using a multilayer Graph-Neural Network model. Our research shares similarities with this study in considering the movements of players within the pitch. However, we differ in both graph topology and purpose. While they focus on predicting outcomes of actions related to gaining possession of the ball, we explore the variation of the importance of specific areas on the pitch across rolling time windows.
In \cite{mclean2018}, a graph-based approach is used to analyze statistical differences among pitch zones in scoring actions, with nodes representing pitch zones obtained by dividing the pitch. Weighted, directed edges denote ball movements between zones during scoring actions, and dominant areas are identified using degree centrality computed for each pitch zone. Our research diverges in several aspects. Firstly, we utilize different types of data sources, with our study relying on GPS data while theirs utilizes video footage. Secondly, when dividing the pitch, we employ methods such as domain expert input or elbow method, whereas their approach is not as well specified. Additionally, our graph topology differs, as we focus on capturing movements and changes in speed zones, whereas their approach uses the number of passes between areas of the pitch. Lastly, our analysis incorporates temporal dynamics, whereas theirs may not.

\section{Methodology}
\label{sec:method}

\begin{figure*}[ht]
    \centering
    \includegraphics[scale=0.45]{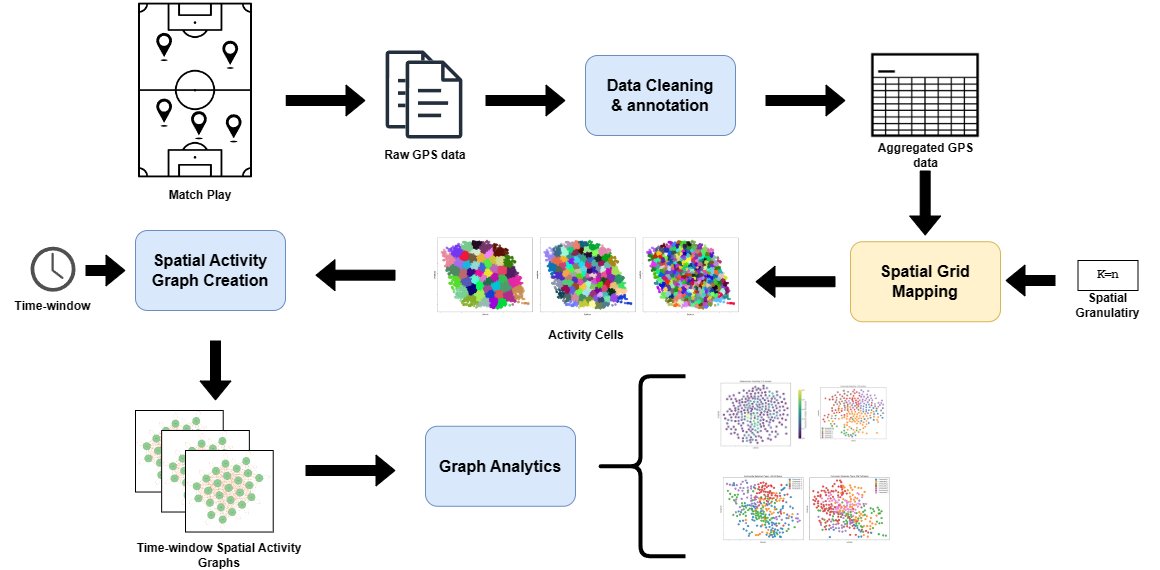}
    \caption{Overview of framework for spatio-temporal graph construction  }
    \label{fig:framework-overview}
\end{figure*}

This work presents a novel framework which provides graph transformations from spatiotemporal datasets of team-based sports. Through this framework we are able to perform graph-based analytics which permits the extraction of insights about players' activity during games.

The framework is composed of four main steps:

\begin{itemize} 
    \item \textbf{Step 1: Data collection, cleaning \& Annotation}. This step involves the initial export of data from the GPS sensing devices, cleaning and annotation of additional features (such as average speed) to the data to produce a dataset of player actions. 
    \item \textbf{Step 2: Spatial Grid Mapping}. This provides an overlay topology for the graph where a grid is developed to allow for more fine-grained spatial analyses. 
    \item \textbf{Step 3: Spatio-Temporal Graphs Construction}. This process builds the spatio-temporal graphs through an iterative process.
    \item \textbf{Step 4: Graph-based Analysis}. This step implements graph theory techniques to analyze patterns of behaviors of players during games or time windows of the same game.  
\end{itemize}

\subsection{Data collection, cleaning \& annotation}

The dataset used is a manipulation of raw GPS data collected during 11 competitive GF inter-county games over the seasons 2019--2021. A micro 10Hz GPS sensor device (STATSports Apex 10 Hz, Northern Ireland, UK) wore by the athletes, recorded 10 observations for each second for the variables latitude, longitude, and speed (m/s). The data have been filtered and pre-processed by the STATSports software (version: 4.5.19). Past research~\cite{beato2018} claimed that the error made by 10 Hz Apex unit (1–2\% of the distances measured in the experiments) can safely be ignored. The data have been transformed into a new dataset by using a series of steps described in a previous work~\cite{antonini2024}. The dataset describes the actions performed by players during games by using some features such as start and end location, average speed, duration of the action, distance covered, etc. A sample of the resulting dataset is shown in Tab.~\ref{tab:actions_dataset} which presents a subset of columns: {\tt Start Lat.} and {\tt Start Lon} represent the initial point of an action,{\tt End Lat} and {\tt End Lon} represent the final point of an action, {\tt Speed (m/s)} is the average speed in metres per second for the action. {\tt Action} is a textual classification of the action (e.g. running, jogging) and {\tt Duration} is the duration in seconds of the action.

\begin{table*}
\caption{Sample of the actions dataset. }\label{tab:actions_dataset}
\resizebox{\textwidth}{!}{
\begin{tabular}{|c|c|c|c|c|c|c|c|c|c|}
\hline
\textbf{Player ID}	& \textbf{Start Time}& \textbf{End Time}& \textbf{Start Lat.}& \textbf{Start Lon.}& \textbf{End Lat.}& \textbf{End Lon.}& \textbf{Speed (m/s)}& \textbf{Action} & \textbf{Duration}\\
\hline
152 & 0 & 4 & 54.123 & $-$7.357 & 54.224 & $-$7.351  & 5.36 &  Running & 4
\\
152 & 3 & 5 & 54.224 & $-$7.351 & 54.011 & $-$7.391  & 3.97 &  Jogging & 2
\\
152 & 5 & 10 & 54.011 &  $-$7.391 & 54.349 & $-$7.650  & 4.98 &  Running & 5
\\
152 & 10 & 16 & 54.349 & $-$7.650 & 54.012 & $-$7.655  & 3.83 &  Jogging & 6
\\
152 & 16 & 20 & 54.012 & $-$7.655  & 54.020 & $-$7.657  & 4.51 &  Running & 4
\\
\hline
\end{tabular}}
\end{table*}

\subsection{Spatial Grid Mapping}
\label{sec:sgm}

In sports analytics, the process of creating a spatial grid overlay for a graph represents a dynamic mapping of the field of play. This process is crucial for certain types of analytics. The segmentation of the field of play and subsequent allocation of the relative coordinates to those areas enables the end user to understand the complex and dynamic movement patterns of players that occur during match play, identify the most active segments of the field of play, and examine the distribution of events. 

The number of unique latitude and longitude coordinates covered by players during games is considerable. These differences can occur for a number of reasons such as subbing players or additional injury time. These factors coupled with the individual dynamics  across games such as offensive and defensive play necessitate examining each game individually constructing a set of cells representing the dynamics of the game.

For each game a spatial grid is constructed by using the Tesspy library~\cite{tesspy}. Tesspy is a library that performs tessellation, the process of dividing space into non-overlapping, gap-free subspaces. The process to create a spatial grid mapping of the pitch is composed of the following steps:

\begin{enumerate}
    \item \textit{Specification of a Point of Interest (POI)}. Tesspy takes as input a POI (example: 'Wembley') and creates a Tesselation object.
    \item \textit{Spatial Grid Generation}. Define an initial resolution for the cell size. Generate adaptive square grids based on the provided Points of Interest (POI) data, starting with this initial resolution. It is returned the set of latitude and longitude coordinates of the centroids of the square grids.
    \item \textit{Spatial Grid Mapping}. Each coordinate covered by players during the game is assigned to the closest centroid. Centroids without assigned coordinates are excluded from the analysis, as they represent areas of the stadium or portions of the pitch not covered by the players.
\end{enumerate}

Each centroid outputted by the tesselation process represent a cell of the pitch which will be used to construct a spatio-temporal graph. 

\subsection{Spatio-Temporal Graphs Construction}
Once the cells are obtained from the spatial grid mapping, a domain expert will provide a time interval (minutes)  to create a set of snapshots of spatial activity within the cell using a rolling window. This produces the \textit{Time-Window Spatial Activity Graph}. 

\paragraph{Definition 1. Time-Window Spatial Activity Graph (TWG)}. The Time-Window Spatial Activity Graph $TWG$ is a set of spatial-activity graphs where each graph is a snapshot of activity for a timepoint $w$ within a cell $a$. Each Spatial Activity Graph is a directed graph $TWG_{a,w} = (N,E)$ where $N$ is the set of nodes representing an area of activity denoted a tuple containing the latitude and longitude for said area $n= <lat, lng>, n\in N$. $E$ denotes the set of edges within the graph where each edge $e$ represents a four-tuple $e = < n_{from}, n_{to},P,\bar{Speed}>, e \in E$ where, $n_{from}$ and $n_{to}$ are the \textit{to} and \textit{from} nodes for the directed edge,  $P$ is the distinct list of players who traversed between two areas of activity and $\bar{Speed}$ is the average speed for all players during that transition. The pseudocode for constructing the \textit{TWG} is presented in Alg 1, where $TW$ is a distinct list of time windows (e.g. $[0,5],[1.6]$...) and $A$ is the set of actions annotated with the cell number.

\begin{figure}[ht]
    \centering
    \includegraphics[scale=0.89]{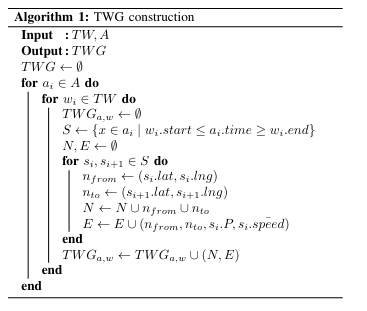}
\end{figure}



\begin{figure*}
    \centering
    \includegraphics[scale=0.4]{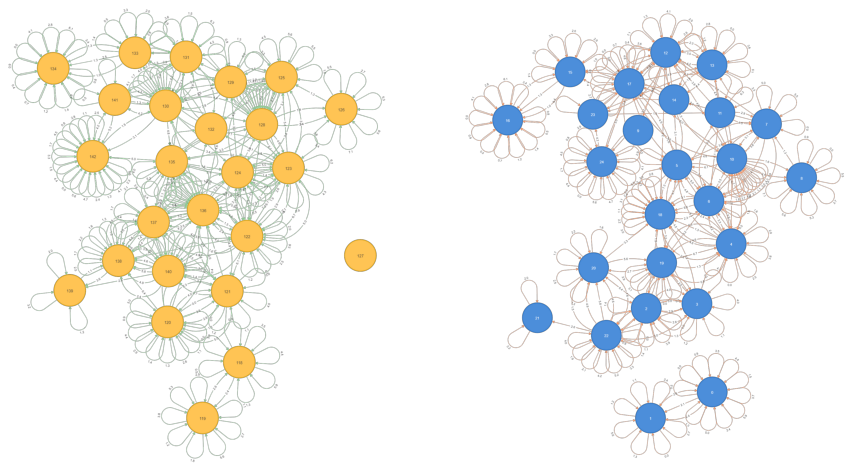}
    \caption{Sample of 25 nodes of the TWG for time windows: [0,5) (\textbf{left}) and [1,6) (\textbf{right}) in a selected game before edges aggregation.}
    \label{fig:graphs_comp}
\end{figure*}

A sample of TWG at different time points can be seen in Fig \ref{fig:graphs_comp}. While the number of nodes given by the spatial grid mapping remains the same, the number of edges may vary (540, and 527 respectively), such as their average weight (3.4, and 3.3 respectively). This results in having more or less connected nodes or to the formation of isolated nodes.

\subsection{Graph-based Analysis}
\label{sec:g-analysis}
Using the TWG generated from the previous step a domain expert can utilise graph analytics to identify areas of activity within the graph, perform comparative analytics across time-windows and identify shifting communities as the game progresses. We now detail two graph analytics methods we can utilize to extract insights into team dynamics. These methods will be demonstrated in our case study in \S\ref{sec:results}. 

\subsection{Nodes Centrality Analysis}
Betweenness centrality detects the degree of influence a node (player or location) maintains over the flow of \emph{information}. 
Nodes that more frequently lie on shortest paths between other nodes will have higher betweenness centrality scores. The formula for calculating betweenness scores \cite{brandes2001} is shown in Equation (\ref{eq:betweennees}) where $\sigma_{st}$is the total number of shortest path going from \textit{s} to \textit{t}, and $\sigma_{st}(x)$ the number of those paths passing through node \textit{x}.  

\begin{equation}
    \centering
    bc(x) = \sum_{s\neq x \neq t}\frac{\sigma_{st}(x)}{\sigma_{st}}
    \label{eq:betweennees}
\end{equation}

In a spatio-temporal analysis such as that used here, the analysis seeks to identify locations through which most shortest paths are routed. A high \emph{betweenness} score may indicate a \emph{bridge} which joins two or more segments of a graph, indicating a location whose removal could effectively segment the graph. However, the most common effect is that all shortest paths become longer, meaning a complete change in the dynamics of multiple player movements. 

Within a TWG, nodes with high weighted betweenness centrality scores indicate pathways frequently traversed by players and represent areas where players transition between speed zones frequently. These pathways are crucial for player movement and can reveal strategic zones on the pitch. Such methods could be employed a sports analyst to plan and optimise player positioning.

\subsection{Community Detection}
This analysis focuses on applying community detection techniques to group areas of the pitch according to their connectivity during a given time window. The communities identified by the Louvain algorithm represent clusters of locations that are frequently connected or have similar movement patterns.
    
A hierarchical clustering approach was used (Louvain modularity) which compares density among the relationships inside the partitions against those outside using Equation(\ref{eq:louvain}), where $A_{i,j}$ represents the weight of the edge between i and j, $k_{i}=\sum_{j}A_{i,}$ is the sum of the weights of the edges for node i, $c_{i}$ is the community to which node $i$ is assigned, and $\delta$ is a function such that $\delta c_{i}c_{j}=1$ if node i and node j are assigned to the same community \cite{blondel2008}. 

\begin{equation}
    \centering
    M = \frac{1}{2m}\sum_{i,j}[A_{i,j}-\frac{k_{i}k_{j}}{2m}]\delta c_{i}c_{j}
    \label{eq:louvain}
\end{equation}

Each community consists of locations that are more closely related in terms of the average speed of movements between them. Community detection can reveal clusters of nodes representing areas where players frequently interact or transition between speed zones together. Analysts can interpret these communities to understand how the team organizes itself spatially and how players collaborate within different areas of the pitch.

\section{Results}
\label{sec:results}
\subsection{Case Study}

This research utilizes data obtained from male Gaelic Football players. GF is a sport originating from Ireland with distinctive rules blending elements of rugby and soccer. In GF, players are permitted to carry the ball in hand but must bounce or tap it every four steps. Passing can be executed using both hands and feet, and scoring can occur above (one point) or below (three points) the crossbar. GF matches involve two teams comprising 15 amateur players each, played on a pitch typically spanning 130–150 meters in length and 80–90 meters in width. Games can be played at the local or county levels, with county matches lasting 70 minutes, divided into two halves of 35 minutes each.

Past research measured that the average match distance is 8160 $\pm$ 1482 meters, with 1731 $\pm$ 659 meters covered at high speed \cite{malone2016}. The sprint distance is 445 $\pm$ 169 meters distributed across 44 sprint actions. The peak speed is 8.4 $\pm$ 0.5 meters/seconds with an average speed of 1.8 $\pm$ 0.3 meters/seconds. 

The proposed graph-based framework has been evaluated using data from a single GF county game to demonstrate its efficacy. 
All players involved in the game have been included in the analysis, except for the goalkeeper.

The match selected as a case study was a county-level male Gaelic Football match with 16 players and ran for 78 minutes resulting in a dataset of 13586 records.

\subsection{Spatial Grid Mapping of the pitch}
\begin{figure*}
    \centering
    \includegraphics[width=\textwidth]{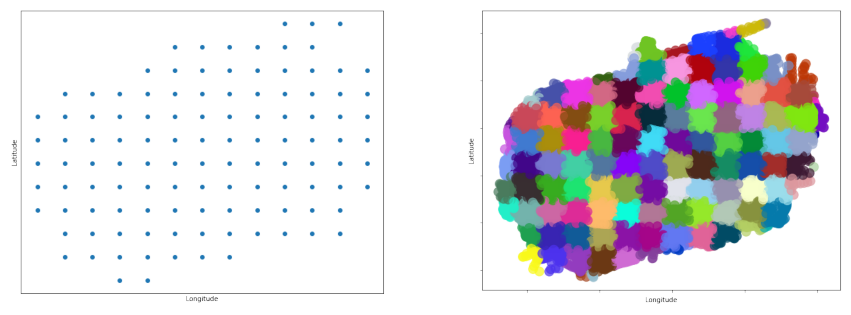}
    \caption{Spatial grid mapping of the pitch. \textbf{Left}: grid points returned by the library. \textbf{Right}: the result of the association of players' coordinates to their closest grid point.}
    \label{pitch_cells}
\end{figure*}

This process starts by specifying the POI (the name of the stadium) and a cell resolution. The tesspy library provides the coordinates of the points representing the pitch as a grid (grid points). Each coordinate covered by players during the game, obtained using GPS data of players, is associated with the closest grid points. The result is shown in Fig.\ref{pitch_cells}

%

At this stage, the original dataset (Tab. \ref{tab:actions_dataset}) is updated by adding the coordinates of the grid points to which the starting and ending locations of the action belong.

\subsection{TWG generation}
Once the spatial grid mapping is complete, the next step is the generation of the TWG. The initial phase involves the definition of the rolling time window. This decision can be made by a domain expert, depending on the granularity of the interest. Larger time windows may lead to more connected graphs but to the loss of information in the aggregating process. On the contrary, narrow time windows can lead to the analysis of few actions, resulting in the generation of limited and less valuable insights.

For the purpose of this paper, a 5-minute time rolling window has been chosen: [0,5), [1,6), etc. [0,5) denotes that the range includes the first value but not the second. This time length permits the study of an elevated number of actions without the loss of significant information due to the edges aggregation. With the game lasting 78 minutes and using a 5-minute rolling window, a TWG of size 74 has been generated. Tab. \ref{tab:graphs_stats} presents descriptive metrics regarding the structure of TWG across a sample of consecutive 5-minute rolling windows. The columns describe the following graphs' characteristics: {\tt Time Window (T.W.)} indicates the interval time used to create the graph. {\tt Nodes} indicates the number of nodes in the graph (cells in the pitch). This number indicates the number of cells representing the segmented pitch (the selected $k$ for the $k$-means algorithm). {\tt Edges} specifies the number of edges (directed movements between cells).{\tt Average Weight Edges} measures the average weight of the edges.{\tt Density} is computed as the ratio of the number of edges in the graph to the maximum possible number of edges and measures the connectivity of the graph. Values are in the range [0,1]. 0 indicates a poorly connected graph, and 1 when all nodes are connected. 
    {\tt Average Degree} is the average number of incoming or outgoing edges per node.
    {\tt Average Clustering Coefficient} measures the average degree to which nodes tend to cluster together. Values are in the range [0,1]. 0 indicates that the nodes are not connected in communities, and there are no triangles formed by connected triples of nodes. 1 indicates that every connected triple of nodes forms a triangle. 
    {\tt  Average Shortest Path} computes the average shortest path length of the graph between all pairs of nodes.

As indicated by the metrics in Tab. \ref{tab:graphs_stats}, the graph's topology varies according to the analyzed window. A lower number of edges may suggest prolonged movements within the same speed zone, indicating less frequent changes in players' speeds. Alternatively, it could indicate a greater concentration of players within specific cells, resulting from the aggregation of edges linking the same nodes during the graph construction. The average weight of the edges describes how the average speed of the players varies across time windows. The density is close to 0 in each time window, meaning the graphs are poorly connected, and may contain isolated nodes. The low connectivity is due to the distance between pitch cells. Typically, players change speed zones after a few seconds, executing movements that cover short distances occurring within the same or adjacent cells. The shortage of edges between cells is also highlighted by the low values of the average clustering coefficient. This results in communities composed of adjacent cells where not all cells within the same community are interconnected.

\begin{table*}
\caption{Descriptive statistics regarding the characteristics of the TWG within each time window.  }\label{tab:graphs_stats}
\resizebox{\textwidth}{!}{
\begin{tabular}{|c|c|c|c|c|c|c|c|}
\hline
\textbf{T.W.} & \textbf{Nodes} & \textbf{Edges} & \textbf{Avg. Weight E.}& \textbf{Density} & \textbf{Avg. Degree} & \textbf{Avg. Clust. Coeff.} & \textbf{Avg. Shortest Path}\\
\hline
$[$0, 5) & 118 & 952&3.2&0.07 & 11.4 & 0.26 &3.5\\
$[$1,6)  & 118 & 882&3.1&0.06 & 10.7 &0.21  &3.8\\
$[$2,7)  & 118 & 872&3.0&0.06 & 10.8 &0.23  &3.8\\
$[$3, 8) & 118 &881 & 2.9&0.05  &10.2&0.20&4.1\\
$[$4, 9) & 118 & 782&2.7&0.05  &9.6&0.19&4.1\\
$[$5, 10)& 118 &718 &2.5&0.05  &8.7&0.18&4.4\\
$[$6, 11)& 118 &764 &2.6&0.05  &9.2&0.20&4.4\\
$[$7, 12)& 118 &726 &2.6& 0.06 &9.1&0.17&4.0\\
$[$8, 13)& 118 &700 &2.6&0.06  &9.3&0.21&3.7\\
$[$9, 14)& 118 &829 &2.8&0.06  &10.0&0.22 & 3.9\\
$[$10, 15)&100 &826 &2.9&0.05  &10.4&0.23&3.8\\
\hline
\end{tabular}}
\end{table*}

\subsection{Analysis of Areas of Activity}
Using betweenness centrality (Sec. \ref{sec:g-analysis}) we can compare the centrality of cells across time windows. As the game evolves betweenness centrality highlights areas of activity as the game progresses. 

\begin{figure*}[ht!]
    \centering
    \includegraphics[width=\textwidth]{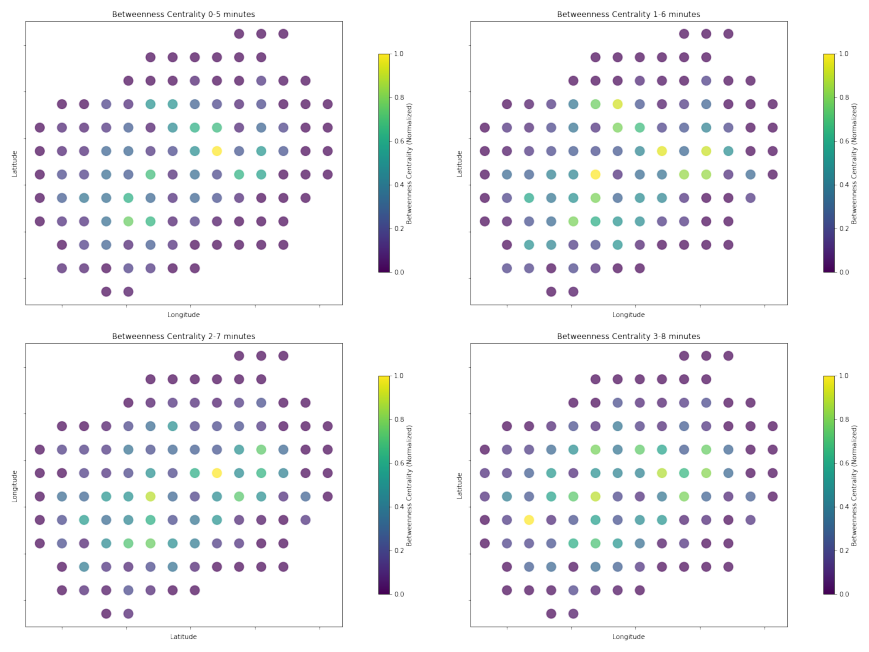}
    \caption{Betweenees centrality for 5 minutes rolling windows:  
    [0, 5) (\textbf{top-left}), [1, 6) (\textbf{top-right}), [2, 7) (\textbf{bottom-left}), and [3, 8) (\textbf{bottom-right}).}
    \label{bet_comp}
\end{figure*}

Fig.~\ref{bet_comp} exhibits the betweenness centrality for four different rolling windows of 5 minutes. The locations displayed are the ones covered by players only in the specific time windows and can change over time. The most important locations (in terms of speed and connectivity) are frequently placed in the central area of the pitch. The dynamic of the game shifts as we move across windows, highlighting strategic alterations in the distribution of the scores. The importance of locations changes during the time windows analyzed: Fig.~\ref{bet_comp} illustrates this transition from central to peripheral areas.
The importance of the cells is given by their connectivity. This means that these cells are reached by a higher number of different cells. Players tend to move from or in these specific cells more than others.

\subsection{Community Detection of Interconnected Areas}
\begin{figure*}[ht!]
    \centering
    \includegraphics[width=\textwidth]{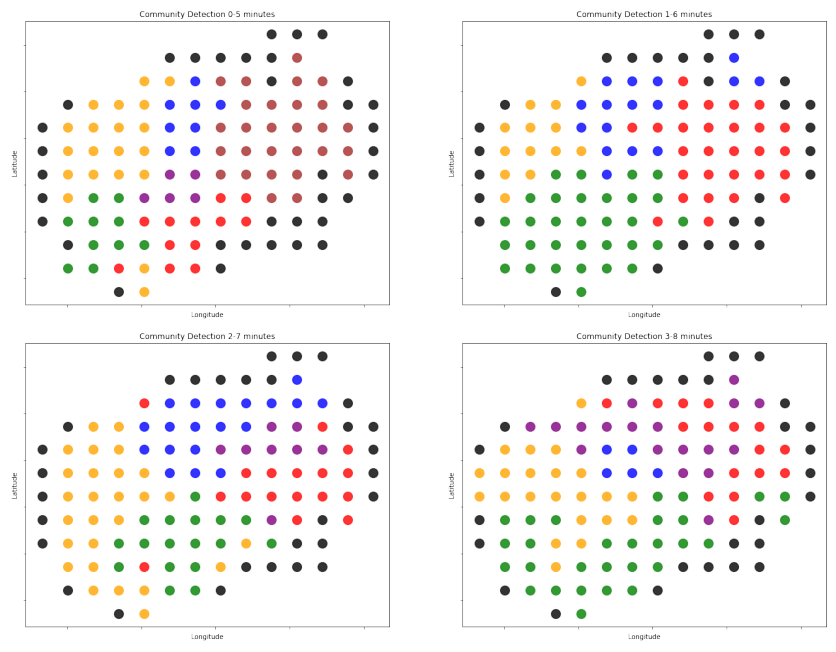}
    \caption{Community detection for 5 minutes rolling windows:  
    [0, 5) (\textbf{top-left}), [1, 6) (\textbf{top-right}), [2, 7) (\textbf{bottom-left}), and [3, 8) (\textbf{bottom-right}). Black points represent coordinates not covered by players in the analyzed time window.}
    \label{fig:com_det_comp}
\end{figure*}

Similar to the fluctuation observed in centrality scores, we employed the Louvain community detection algorithm to identify distinct communities across different time windows in the TWG. Fig. \ref{fig:com_det_comp} details the community detection result across four consecutive time windows (black points represent coordinates not covered by players in the time window). 
The algorithm returns a variable number of communities for each time window (6, 4, 5, and 5 respectively). These communities change shapes and internal composition over time. This behaviour may derived from a wider spread of players across the pitch, leading to a greater number of less populated communities. The detection of communities formed mainly by adjacent cells suggests that players tend to traverse short distances before changing speed zones. In contrast, when a cell belongs to a different community compared to its adjacency cells means that the specific cell has been visited by players covering different areas, potentially reflecting the initiation of actions in remote areas. This observation provides insights into how players navigate the pitch. In particular, where they tend to change speed zone. It provides valuable insights into players' strategies and preferences, which can be particularly useful for a sports coach. 

The internal composition of the identified communities, in terms of descriptive statistics regarding the speed originating from nodes within those communities, remains relatively stable. The communities are formed mainly thanks to the proximity of the nodes, which makes it easier and more frequent to perform a movement in the same community. Within the [0,5) time frame, community red stands out with notably higher metrics for the variable speed. It exhibits the highest recorded values for the 1$^{st}$ quartile (2.4 m/s), average (3.6 m/s), and 3$^{rd}$ quartile (4.9 m/s), distinguishing itself from other communities in this time window. In the same area of the pitch, in the successive time window [1,6) we observe a similar pattern: community green records the highest values for 1$^{st}$ quartile (1.6 m/s), average (3.2 m/s), and 3$^{rd}$ quartile (4.6 m/s).  Likewise, community green in window [2, 7) and community yellow in window [3, 8) demonstrate similar trends. This suggests an area characterized by a greater frequency of activities occurring at higher speeds.

In contrast, it seems that communities colored yellow and brown in the time window [0,5), yellow in [1,6), yellow and red in [2,7), and red in [3,8) indicate areas where activities typically occur at slower speeds. For instance, in the yellow community (which covers roughly the same area) during the time windows [0,5), [1,6), and [2,7), the recorded speeds for the 1$^{st}$ quartile are 1.6, 1.6, and 1.5 m/s respectively. The averages are 2.9, 2.8, and 2.7 m/s, and the 3$^{rd}$ quartiles are 4.2, 3.9, and 3.9 m/s. These values represent some of the lowest scores observed for these metrics.

\section{Conclusion \& Future work}
\label{sec:conc}
As the domain sports analytics evolves and more field-based teams and sporting organisations adopt data-driven approaches to training and strategy, emphasis on analytics will shift from individualised player performance to team-focused analytics necessitating an investment in data engineering solutions to transform, clean and structure the data in a format suitable for extracting team-centred analytics during match play. In this work we present a framework to construct a Time Window Spatial Activity Graph; a collection of spatio-temporal graphs across rolling time windows focused on player movement during match play. 
Our evaluation demonstrates how graph-analytics such as betweenness centrality and community detection can be utilised to extract insights into team-dynamics and may aid sports analysts in areas of player positioning and team strategies. Our future work is to employ the use of more graph-analytics methods such as link-prediction to predict graph state changes across time windows.

\section*{Acknowledgment}
This work was supported by Science Foundation Ireland through the Insight Centre for Data Analytics (SFI/12/RC/2289\_P2) and the SFI Centre for Research Training in Machine Learning (18/CRT/6183).

\end{document}